\documentclass[aps,prd,twocolumn,nofootinbib,showpacs]{revtex4}

\usepackage{graphicx,color}
\usepackage[colorlinks=true,linkcolor=blue,citecolor=blue,urlcolor=blue]{hyperref}

\begin{document}

\title{Predictions for the Top-Quark Forward-Backward Asymmetry at High Invariant Pair Mass Using the Principle of Maximum Conformality}

\author{Sheng-Quan Wang$^{1,2}$}
\email[email:]{sqwang@cqu.edu.cn}

\author{Xing-Gang Wu$^1$}
\email[email:]{wuxg@cqu.edu.cn}

\author{Zong-Guo Si$^3$}
\email[email:]{zgsi@sdu.edu.cn}

\author{Stanley J. Brodsky$^4$}
\email[email:]{sjbth@slac.stanford.edu}

\address{$^1$Department of Physics, Chongqing University, Chongqing 401331, P.R. China}
\address{$^2$School of Science, Guizhou Minzu University, Guiyang 550025, P.R. China}
\address{$^3$Department of Physics, Shandong University, Jinan, Shandong 250100, P.R. China}
\address{$^4$SLAC National Accelerator Laboratory, Stanford University, Stanford, California 94039, USA}

\date{\today}

\begin{abstract}

The D0 collaboration at FermiLab has recently measured the top-quark pair forward-backward asymmetry in $\bar p p \to t \bar{t} X$ reactions as a function of the $t\bar{t} $ invariant mass $M_{t\bar{t}}$. The D0 result for $A_{\rm FB}(M_{t\bar{t}}>650\; {\rm GeV})$ is smaller than $A_{\rm FB}(M_{t\bar{t}})$ obtained for small values of $M_{t\bar{t}}$, which may indicate an ``increasing-decreasing" behavior for $A_{\rm FB}(M_{t\bar{t}}>M_{\rm cut})$. This behavior is not explained using conventional renormalization scale-setting, even by a next-to-next-to-leading order (N$^2$LO) QCD calculation -- one predicts a monotonically increasing behavior.

In the conventional scale-setting method, one simply guesses a single renormalization scale $\mu_r$ for the argument of the QCD running coupling and then varies it over an arbitrary range. However, the conventional method has inherent difficulties. For example, the resulting pQCD predictions depend on the choice of renormalization scheme, in contradiction to the principle of ``renormalization scheme invariance'' -- predictions for physical observables cannot depend on a theoretical convention. The error estimate obtained by varying $\mu_r$ is unreliable since it is only sensitive to perturbative contributions involving the pQCD $\beta$-function. Worse, guessing the renormalization scale gives predictions for precision QED observables which are in contradiction to results obtained using the standard Gell-Mann-Low method. In contrast, if one fixes the scale using the Principle of Maximum Conformality (PMC), the resulting pQCD predictions are renormalization-scheme independent since all of the scheme-dependent $\{\beta_i\}$-terms in the QCD perturbative series are resummed into the QCD running couplings at each order. The $\{\beta_i\}$-terms at each order can be unambiguously identified using renormalization group methods such as the $R_\delta$ method. The PMC then determines the renormalization scales of the running coupling at each order and provides unambiguous scale-fixed and scheme-independent predictions. The PMC reduces in the $N_C\to 0$ Abelian limit to the standard Gell-Mann-Low scale-setting method used in QED, including precise scheme-independent predictions for the forward-backward symmetry of the $e^+ e^- \to \mu^+ \mu^-$ cross-section.

By using the rigorous PMC scale-setting procedure, one obtains a comprehensive, self-consistent pQCD explanation for the Tevatron measurements of the top-quark pair forward-backward asymmetry. In this paper we show that if one applies the PMC to determine the top versus anti-top quark forward-backward asymmetry by properly using the pQCD predictions up to N$^2$LO level, one obtains the predictions without renormalization scheme or scale ambiguities. For example, the PMC predicts $A^{\rm PMC}_{\rm FB}(M_{t\bar{t}}>450~\rm GeV)=29.9\%$ at the Tevatron, which is  consistent with the CDF measurements. In addition, the PMC prediction for $A_{\rm FB}(M_{t\bar{t}}> M_{\rm cut})$ shows an ``increasing-decreasing" behavior for increasing values of $M_{\rm cut}$ which is not observed in the NLO and N$^2$LO predictions for $A_{\rm FB}(M_{t\bar{t}}> M_{\rm cut})$ with conventional scale-setting. This behavior could be tested by the future more precise measurements at the LHC.


\end{abstract}

\maketitle

\section{Introduction}

Measurements of the top versus anti-top quark asymmetry in $\bar p p \to t \bar t X$ reactions at the Tevatron have provided important tests of perturbative quantum chromodynamics (pQCD) and the Standard Model. The forward-backward $t \bar t$ asymmetry is defined as
\begin{eqnarray}
A_{\rm FB} &=&\frac{N(\Delta y>0)-N(\Delta y<0)} {N(\Delta y>0)+N(\Delta y<0)},
\end{eqnarray}
where $\Delta y=y_{t}-y_{\bar{t}}$ is the difference between the rapidities of top and anti-top quarks, and $N$ stands for the number of events. This asymmetry is due in QCD to the interference of perturbative amplitudes with different charge conjugation, such as the one-gluon and two-gluon annihilation contributions to the $q \bar q \to t \bar t$ subprocess.

A review of the theoretical and experimental features of $A_{\rm FB}$ can be found in Ref.\cite{Aguilar-Saavedra:2014kpa}. The initial Standard Model (SM) predictions for $A_{\rm FB}$, which were based on pQCD at next-to-leading order (NLO)~\cite{Kuhn:1998jr, Kuhn:1998kw}, appeared to be in substantial disagreement with the Tevatron CDF and D0 measurements~\cite{Aaltonen:2008hc, Aaltonen:2011kc, Aaltonen:2012it, Abazov:2007ab, Abazov:2011rq, Abazov:2014cca}. The SM predictions for the $t \bar t$ asymmetry $A_{\rm FB}$ were subsequently improved by including electroweak contributions: $(8.9^{+0.8}_{-0.6})\%$~\cite{Hollik:2011ps}, $(8.7\pm1.0)\%$~\cite{Kuhn:2011ri}, and $(8.8\pm0.6)\%$~\cite{Bernreuther:2012sx}. More recently, a full next-to-next-to-leading order (N$^2$LO) pQCD calculation was performed; it predicts $A_{\rm FB}=(9.5\pm0.7)\%$~\cite{Czakon:2014xsa}. Ref.~\cite{Kidonakis:2015ona} has found that the soft-gluon corrections are important to the asymmetry, and by including those corrections via a proper resummation approach, the author provides an approximate N$^3$LO QCD prediction to the asymmetry, i.e. $A_{\rm FB} = (10.0\pm0.6)\%$. These N$^2$LO and higher order predictions agree with the Tevatron measurement within errors, $A^{\rm D0}_{\rm FB}=(10.6\pm3.0)\%$~\cite{Abazov:2014cca} and $A^{\rm D0}_{\rm FB}=(11.8\pm2.5\pm1.3)\%$~\cite{Abazov:2015fna} in D0 collaboration and $A^{\rm CDF}_{\rm FB}=(16.4\pm4.7)\%$~\cite{Aaltonen:2012it} and $A^{\rm CDF}_{\rm FB}=(12\pm13)\%$~\cite{Aaltonen:CDF} in CDF collaboration.

The NLO pQCD predictions with conventional scale-setting for $A_{\rm FB}(M_{t\bar{t}} > 450 \;{\rm GeV})$ are $(12.8^{+1.1}_{-0.9})\%$~\cite{Hollik:2011ps}, $(12.8\pm 1.1)\%$~\cite{Kuhn:2011ri}, and $(12.9^{+0.8}_{-0.6}) \%$~\cite{Bernreuther:2012sx}, which deviate from the 2011 CDF measurement $(47.5\pm11.4)\%$~\cite{Aaltonen:2011kc} by about $3.4\sigma$ standard deviations and the 2013 CDF measurement $(29.5\pm5.8\pm3.3)\%$~\cite{Aaltonen:2012it} by about $2.4\sigma$ standard deviation. While an update of $A_{\rm FB}(M_{t\bar{t}}>450\;{\rm GeV})$ using the N$^2$LO and approximate N$^3$LO calculations with conventional scale-setting is not yet available, a measurement of the differential $A_{\rm FB}(M_{t\bar{t}})$ has been provided by both CDF and D0 Collaborations~\cite{Aaltonen:2012it, Abazov:2014cca} and compared with the N$^2$LO and N$^3$LO results with conventional scale-setting~\cite{Czakon:2014xsa, Kidonakis:2015ona}. Agreement was found with the D0 data but not with the CDF data at large values of $M_{t\bar{t}}$. Moreover, the new D0 result for $A_{\rm FB}(M_{t\bar{t}}>650\; {\rm GeV})=-12.3\pm29.6$~\cite{Abazov:2014cca}, which is smaller than $A_{\rm FB}(M_{t\bar{t}})$ obtained for small values of $M_{t\bar{t}}$, may suggest an ``increasing-decreasing" behavior for $A_{\rm FB}(M_{t\bar{t}}>M_{\rm cut})$. This ``increasing-decreasing" behavior was not reflected in the pQCD predictions with conventional scale-setting. It is interesting to show whether one can achieve such an ``increasing-decreasing" behavior by using the PMC scale-setting.

All of the SM predictions discussed above have been based on conventional scale-setting; i.e., one simply takes the renormalization scale as the top-quark mass $m_{t}$ and then varies the scale over an arbitrary range, i.e. $[m_{t}/2, 2m_{t}]$, in order to estimate the scale uncertainty. However, this procedure of guessing the renormalization scale, although conventional, gives renormalization scheme-dependent predictions. It also leads to a non-convergent renormalon perturbative series. Moreover, one would obtain incorrect results if one applies this method to QED processes~\cite{Wu:2013ei}. It is possible for conventional scale-setting to accidentally predict the correct value of a global observable such as the total cross-section at sufficiently high order; however, since one assumes the same renormalization scale at each order in $\alpha_s$, it will often give incorrect predictions for each perturbative order correction. In fact, the renormalization scale and effective number of flavors are in general distinct at each order of pQCD, reflecting the different virtualities of the subprocesses as a function of phase-space. This provides the underlying reason why a single `guessed' scale cannot explain the ``increasing-decreasing" behavior of $A_{\rm FB}$ as the $t \bar t$-pair mass is varied.

In contrast, the Principle of Maximum Conformality (PMC)~\cite{Brodsky:2011ta, Brodsky:2012rj, Brodsky:2011ig, Mojaza:2012mf, Brodsky:2013vpa} provides a systematic way to eliminate the renormalization scheme-and-scale uncertainties. The PMC reduces in the $(N_c\to 0)$ Abelian limit~\cite{Brodsky:1997jk} to the standard Gell-Mann-Low method~\cite{GellMann:1954fq}, where all vacuum polarization contributions are associated with the dressed photon propagator and are thus resummed to determine the optimal scale of the running coupling. The PMC provides the underlying principle for the Brodsky-Lepage-Mackenzie approach~\cite{Brodsky:1982gc}, extending it unambiguously to all orders using renormalization group methods.

The authors of Ref.\cite{Czakon:2014xsa} have noted that an alternative scale-setting procedure, called the ``large $\beta_0$-approximation"~\cite{Neubert:1994vb, Beneke:1994qe}, leads to incorrect pQCD $n^2_f$-terms at NNLO. It should be emphasized that this analytic error is a defect of the ``large $\beta_0$-approximation"; it does not occur if one uses PMC scale-setting~\cite{Wu:2013ei}, i.e. the correct $n_f$-series at each perturbative order can be determined via its one-to-one correspondence to the $\beta$-series at the same order.

The PMC has a solid theoretical foundation, satisfying renormalization group invariance~\cite{Wu:2014iba} and all the other self-consistency conditions derived from the renormalization group~\cite{Brodsky:2012ms}. The PMC scales at each order are formed by shifting the arguments of the running coupling to eliminate all non-conformal $\{\beta_i\}$-terms. This elimination is done by using the $\beta$-pattern determined by renormalization group equations; one then obtains correct behavior of the running coupling at each order and at each phase-space point. A systematic method for identifying the $\{\beta_i\}$-terms is given in Refs.~\cite{Mojaza:2012mf, Brodsky:2013vpa}. This procedure also predicts some of the previously unknown higher-order $\{\beta_i\}$-terms via the degeneracy patterns appearing at different orders~\cite{Bi:2015wea}. After applying the PMC, the divergent renormalon series such as $\sum_{n} n! \beta_0^n \alpha_s^n$ does not appear and thus the pQCD convergence is automatically improved. The PMC has been successfully applied to many high-energy processes, including processes which involve multiple physical scales, such as hadronic $Z$ decays~\cite{Wang:2014aqa} and  $\Upsilon(1S)$ leptonic decays~\cite{Shen:2015cta}.

In this paper we will apply the PMC procedure to the calculation of the $A_{\rm FB}$ asymmetry by applying the pQCD results up to N$^2$LO level with the goal of achieving precise pQCD predictions without renormalization scale ambiguities. We shall show that the PMC provides a self-consistent explanation of the recently reported D0 and CDF measurements. In our numerical calculations, we will assume the top-quark mass is $m_{t}=173.1$ GeV and will utilize the CTEQ6.6M parton distribution functions~\cite{Nadolsky:2008zw}. The Bernreuther-Si program~\cite{Bernreuther:2012sx} will be used for evaluating the electroweak corrections, and the HATHOR program will be adopted for treating the two-loop QCD corrections to the total cross-sections~\cite{Aliev:2010zk} \footnote{The HATHOR approximate NNLO-terms have been replaced by the exact NNLO-terms from the Top++ program~\cite{Czakon:2011xx}. }. Thus, the LO and NLO PMC scales can be unambiguity determined by the higher information in HATHOR program. We shall show that the pQCD convergence for the $t\bar{t}$ production cross sections are greatly improved; our PMC predictions, even at low orders, is in good agreement with the complete NNLO prediction~\cite{Czakon:2014xsa}, and provide a self-consistent pQCD explanation for the Tevatron measurements.

\begin{widetext}
\begin{center}
\begin{table}[htb]
\begin{tabular}{|c|c|c|c|c|c|c|c|c|}
\hline
& \multicolumn{4}{c|}{Conventional scale-setting} & \multicolumn{4}{c|}{PMC scale-setting} \\
\hline
~~~ ~~~    &~~~LO~~~  &~~~NLO~~~  &~~~N$^2$LO~~~ &~~~ {\it Total} ~~~&~~~LO~~~  &~~~NLO~~~  &~~~N$^2$LO~~~ &~~~ {\it Total} ~~~\\
\hline
$(q\bar{q})$-channel & 4.901 & 0.960 & 0.481 & 6.346 & 4.760 & 1.728 & -0.0621 & 6.383 \\
\hline
$(gg)$-channel    & 0.542 & 0.434 & 0.156 & 1.132 & 0.540 & 0.516  & 0.149  & 1.230 \\
\hline
$(gq)$-channel    & 0.000 &-0.0361 & 0.0051& -0.0309 & 0.000 & -0.0361 & 0.0051  & -0.0309 \\
\hline
$(g\bar{q})$-channel & 0.000 &-0.0361 & 0.0051& -0.0310 & 0.000 & -0.0361 & 0.0051  & -0.0310 \\
\hline
sum     & 5.444  & 1.322 & 0.647 & 7.416 & 5.300 & 2.172  &  0.0971 & 7.552 \\
\hline
\end{tabular}
\caption{The top-quark pair production cross-sections (in unit: pb) before and after PMC scale-setting at the Tevatron with the collision energy $\sqrt{S}=1.96$ TeV. The initial renormalization scale and the factorization scale are taken as $\mu_r=\mu_f=m_t$.}
\label{tab_totcs}
\end{table}
\end{center}
\end{widetext}

\section{Comparisons of PMC and Conventional Scale-Setting Predictions for the $t \bar t$  production cross-section }

We will first compare the pQCD predictions up to the N$^2$LO level for the total top-quark pair production cross-sections for the Tevatron collision energy $\sqrt{s}=1.96$ TeV using the PMC and conventional scale-settings, respectively. The results are presented in Table \ref{tab_totcs}, where the cross-sections for the individual $(q\bar{q})$-, $(gq)$-, $(g\bar{q})$- and $(gg)$- channels are also presented. The initial scale $\mu_r$ is set as $m_t$ and then varied over the range $\mu_{r}\in[m_t/2, 2m_t]$. The total cross-sections from all the production channels, before and after PMC scale-setting, are
\begin{eqnarray}
\sigma_{\rm Total}|_{\rm Conv.} &=& 7.42^{+0.25}_{-0.29} \; {\rm pb}, \\
\sigma_{\rm Total}|_{\rm PMC}   &\simeq& 7.55 \; {\rm pb}.
\end{eqnarray}
Both the PMC and conventional scale-setting procedures agree with the CDF and D0 measurements within errors~\cite{Aaltonen:2010ic, Abazov:2011mi, Abazov:2011cq, Aaltonen:2013wca}; the recent combined cross-section given by the CDF and D0 collaborations is $7.60\pm0.41$ pb~\cite{Aaltonen:2013wca}. The PMC total cross-section is almost unchanged when one varies the starting scale $\mu_r$. The dependence of the total cross-section on the choice of renormalization scale is also small using conventional scale-setting if one incorporates N$^2$LO QCD corrections.

However, using a single guessed scale does not predict the cross-sections for individual channels correctly at each perturbative order. In fact, by analyzing the pQCD series in detail, we find that the errors for the separate cross-sections at each perturbative order from conventional scale-setting are large in all of the contributing channels.

\begin{table}[htb]
\centering
\begin{tabular}{|c|c|c|c|c|c|c|}
\hline
& \multicolumn{3}{c|}{Conventional} & \multicolumn{3}{c|}{PMC} \\
\hline
~$\mu_r$~  & ~$m_t/2$~  & ~$m_t$~ & ~$2m_t$~ & ~$m_t/2$~ & ~$m_t$~ & ~$2m_t$~  \\
\hline
~$\sigma^{\rm{LO}}_{q\bar{q}}$~ & ~5.989~ & ~4.901~ & ~4.089~ & ~4.759~ & ~4.760~ & ~4.761~ \\
\hline
~$\sigma^{\rm{NLO}}_{q\bar{q}}$~ & ~0.090~ & ~0.960~ & ~1.412~ & ~1.728~ & ~1.728~ & ~1.729~ \\
\hline
~$\sigma^{\rm{N^2LO}}_{q\bar{q}}$~ & ~0.454~ & ~0.481~ & ~0.629~ & ~-0.062~ & ~-0.062~ & ~-0.062~ \\
\hline
\end{tabular}
\caption{The $(q\bar{q})$-channel cross-sections (in unit: pb) at each perturbative order under the conventional and PMC scale-settings, where three typical renormalization scales $\mu_{r}=m_t/2$, $m_t$ and $2m_t$ are adopted. The factorization scale is taken as $\mu_f=m_t$. }
\label{tevat}
\end{table}

As an example, the contributions of the dominant $(q\bar{q})$-channel with and without PMC scale-setting are presented in Table \ref{tevat}. This subprocess is the dominant source of the $t\bar t$ asymmetry. In agreement with expectations, the scale dependence of the total $(q\bar{q})$- cross-section up to N$^2$LO level is small;  i.e., $\Delta\sigma^{\rm Total}_{q\bar{q}}= \sum^{\rm N^2LO}\limits_{i={\rm LO}} \Delta \sigma^{i}_{q\bar{q}} \simeq \pm3\%$ for $\mu_{r}\in[m_t/2,2m_t]$.
For clarity, we define a ratio $\kappa_i$ to illuminate the scale  dependence of individual cross-sections $\sigma^{i}_{q\bar{q}}$ at each order:
\begin{displaymath}
\kappa_i=\frac{\left. \sigma^i_{q\bar{q}}\right|_{\mu_{r}=m_t/2} -\left. \sigma^i_{q\bar{q}}\right|_{\mu_{r}=2m_t}}{\left.\sigma^i_{q\bar{q}} \right|_{\mu_{r}=m_t}},
\end{displaymath}
where $i$=LO, NLO and N$^2$LO, respectively. Using conventional scale-setting, we obtain
\begin{displaymath}
\kappa_{\rm LO}=39\%,~\kappa_{\rm NLO}=-138\%,~\kappa_{\rm N^2LO}=-36\%.
\end{displaymath}
These results show that if one uses conventional scale-setting, then the dependence on the choice of initial scale at each order is very large. For example, the scale dependence of $\sigma^{\rm NLO}_{q\bar{q}}$, which gives the dominant component of the asymmetry $A_{\rm FB}$, reaches up to $-138\%$. One thus cannot determine  accurate values for the individual cross-sections $\sigma^{i}_{q\bar{q}}$ using conventional scale-setting. Under conventional scale-setting, $\sigma^{\rm NLO}_{q\bar{q}}$ increases and $\sigma^{\rm Total}_{q\bar{q}}$ decreases with the increment of $\mu_r$. Thus in order to agree with the measured total cross-section, it prefers a smaller scale $\simeq m_t/2$, which however leads to the prediction of a small $t \bar t$ asymmetry, well below the data. On the other hand, Table \ref{tevat} shows that all of the $\kappa_i$-values become less than $0.1\%$ if one uses the PMC. Thus the scale errors for both the total cross-section and the individual cross-sections at each order are simultaneously eliminated using the PMC scale-setting, and the residual scale dependence due to unknown higher-order $\{\beta_i\}$-terms become negligible~\cite{Brodsky:2012sz}.

\section{Predictions for the $t\bar{t}$ Forward-Backward Asymmetry}

The top-quark pair forward-backward asymmetry in $\bar p p \to t \bar t X$ collisions is also sensitive to the renormalization scale-setting procedure. As shown in Table \ref{tab_totcs} the $(q\bar{q})$-channel provides the dominant contribution to $A_{\rm FB}$ at the Tevatron. It is important to note the NLO PMC scale $\mu^{\rm PMC, NLO}_r$ of the $(q\bar{q})$-channel is much smaller than $m_{t}$ (as shown by Table~\ref{tab_mtcut}). This reflects the fact that the two virtual gluons in the $s$-channel in the annihilation amplitude $q \bar q \to g^*  g^* \to t \bar t$ share the virtuality of the subprocess. The resulting NLO $(q\bar{q})$ cross-section is in fact about twice as large as the cross-section predicted by conventional scale-setting; the precision of the predicted asymmetry $A_{\rm FB}$ is also greatly improved. Moreover, after applying the PMC, the predicted ratio of the cross-section at the N$^2$LO level to the NLO cross-section for the $q\bar{q}$-channel, i.e. $|\sigma_{q\bar{q}}^{\rm N^2LO} / \sigma_{q\bar{q}}^{\rm NLO}|$, is reduced from $\sim 50\%$ to be less than $\sim 4\%$. This indicates a great improvement of pQCD convergence can be achieved by using the PMC. Such an improvement of the pQCD convergence is essential for achieving accurate pQCD predictions for the $t \bar t$ asymmetry. If we further include the $\mathcal{O}(\alpha^2_s\alpha)$ and the $\mathcal{O}(\alpha^2)$ electroweak contributions, we achieve a precise SM ``NLO-asymmetry"  predictions~\cite{Brodsky:2012ik},
\begin{eqnarray} \label{asymmetry}
A^{\rm PMC}_{\rm FB}={\alpha^3_s N_1+\alpha^2_s \alpha \tilde{N}_1 + \alpha^2 \tilde{N}_0 \over \alpha^2_s D_0 + \alpha^3_s D_1},
\end{eqnarray}
where the $D_i$-terms stand for the total cross-sections at each $\alpha_s$-order and the $N_i$-terms stand for the corresponding asymmetric contributions. The term labeled $\tilde{N}_1$ corresponds to the QCD-QED interference contribution at the order ${\cal O}(\alpha^2_s \alpha)$, and $\tilde{N}_0$ stands for the pure electroweak antisymmetric ${\cal O}(\alpha^2)$  contribution  arising from $|{\cal M}_{q\bar{q}\to\gamma\to t\bar{t}}+{\cal M}_{q\bar{q}\to Z^0\to t\bar{t}}|^2$~\cite{Hollik:2011ps}.

Under the conventional scale-setting, the N$^2$LO $N_2$-term provides large contribution for $A_{\rm FB}$~\cite{Czakon:2014xsa}. Given that after applying the PMC scale-setting, the pQCD convergence is greatly improved and the $D_2$-term is lowered by one order of magnitude in comparison to the $D_1$-term, we assume the same behavior for $N_2$ and consider the asymmetric NLO $N_1$-term to provide the dominant contribution. We therefore neglected the asymmetric N$^2$LO $N_2$-term when using the PMC.

Furthermore, to compare the PMC prediction with the asymmetry assuming conventional scale-setting, we further rewrite the PMC asymmetry as~\cite{Brodsky:2012ik}
\begin{eqnarray}
A_{\rm FB}^{\rm PMC} &=&  \left\{\frac{\sigma^{\rm Conv.,LO}_{\rm tot}} {\sigma^{\rm PMC,NLO}_{\rm tot}} \right\} \left\{ \frac{{\overline{\alpha}^3_s}\left(\overline{\mu}^{\rm PMC, NLO}_r\right)} {{\alpha^3_s} \left(\mu^{\rm Conv.}_r\right)} \left. A^{\rm (Conv.)}_{\rm FB}\right|_{\alpha^3_s} + \right.\nonumber\\
&&
\left. \frac{{\overline{\alpha}^2_s}\left(\overline{\mu}^{\rm PMC, NLO}_r\right)} {{\alpha^2_s} \left(\mu^{\rm Conv.}_r\right)} \left. A^{\rm (Conv.)}_{\rm FB}\right|_{\alpha^2_s \alpha}+ \left. A^{\rm (Conv.)}_{\rm FB}\right|_{\alpha^2} \right\}, \label{pmcasy}
\end{eqnarray}
where the symbol ``Conv." stands for the prediction calculated by using the conventional scale-setting, and ``PMC" stands for the corresponding value after applying the PMC. By using Eq.(\ref{pmcasy}), we obtain precise prediction for $A_{\rm FB}$ without renormalization scale uncertainty: $A^{\rm PMC}_{\rm FB} = 9.2\%$.

\section{Predictions for the $t\bar{t}$ Forward-Backward Asymmetry as a Function of the Pair Mass}

\begin{figure}[htb]
\includegraphics[width=0.48\textwidth]{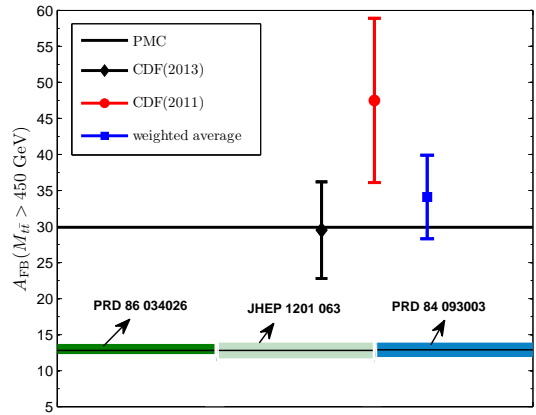}
\caption{Comparison of the PMC prediction for the top-pair asymmetry $A_{\rm FB}(M_{t\bar{t}}>450~\rm GeV)$ with the CDF measurement~\cite{Aaltonen:2011kc, Aaltonen:2012it}. The NLO results predicted by Refs.~\cite{Hollik:2011ps, Kuhn:2011ri, Bernreuther:2012sx} under conventional scale-setting are presented as a comparison, which are shown by shaded bands. } \label{fpbcdfpmc}
\end{figure}

We will next discuss predictions for the top-quark pair asymmetry $A_{\rm FB}(M_{t\bar{t}}>M_{\rm cut})$ as a function of the top-pair invariant mass lower limit $M_{\rm cut}$. In the case of $M_{\rm cut}=450$ GeV, the predicted asymmetry using conventional scale-setting is $A_{\rm FB}(M_{t\bar{t}}>450~\rm GeV)|_{\rm Conv.}=12.9\%$, consistent with previous SM predictions~\cite{Hollik:2011ps, Kuhn:2011ri, Bernreuther:2012sx}. However, after applying the PMC, the predicted asymmetry is much larger: $A_{\rm FB}(M_{t\bar{t}}>450~\rm GeV)|_{\rm PMC}=29.9\%$. A comparison of the PMC prediction with the CDF measurements is presented in Fig.(\ref{fpbcdfpmc}). The PMC prediction agrees with the weighted average of the CDF measurements within errors~\cite{Aaltonen:2011kc, Aaltonen:2012it}. Thus after applying the PMC, the large discrepancy between the SM estimates and CDF measurement predicted using conventional scale-setting is removed.

\begin{widetext}
\begin{center}
\begin{table}[tb]
\begin{tabular}{|c|c|c|c|c|c|c|c|c|c|c|}
\hline
\multicolumn{11}{|c|}{Top-quark pair asymmetries $A_{\rm FB}(M_{t\bar{t}}>M_{\rm cut})$} \\
\hline
 $M_{\rm cut}$(GeV) & ~~~350~~~ & ~~~400~~~ & ~~~450~~~ & ~~~500~~~ & ~~~550~~~ & ~~~600~~~ & ~~~650~~~ & ~~~700~~~ & ~~~750~~~ & ~~800~~ \\
\hline
$A_{\rm FB}(M_{t\bar{t}}>M_{\rm cut})|{\rm Conv.}$ & 8.9\% & 10.9\% & 12.9\% & 14.7\% & 16.4\% & 17.8\% & 19.3\% & 20.5\% & 21.9\% & 23.2\%  \\
\hline
$A_{\rm FB}(M_{t\bar{t}}>M_{\rm cut})|{\rm PMC}$ & 9.6\% & 17.1\% & 29.9\% & 43.5\% & 45.1\% & 37.8\% & 33.5\% & 31.4\% & 30.5\% & 30.1\% \\
\hline
${\overline{\alpha}_s}(\overline{\mu}^{\rm PMC}_r)$ & 0.123 & 0.131 & 0.146 & 0.157 & 0.153 & 0.138 & 0.129 & 0.123 & 0.120 & 0.117 \\
\hline
$\overline{\mu}^{\rm PMC}_r$(GeV) & 71 & 48 & 26 & 18 & 20 & 35 & 53 & 69 & 83 & 94 \\
\hline
\end{tabular}
\caption{Top-quark pair asymmetries $A_{\rm FB}(M_{t\bar{t}}>M_{\rm cut})$ using conventional (Conv.) and PMC scale-setting procedures, respectively. The Conv. predictions are for the NLO pQCD predictions with $\mathcal{O}(\alpha^2_s\alpha)$ and the $\mathcal{O}(\alpha^2)$ electroweak contributions and the PMC predictions are calculated by Eq.(\ref{pmcasy}). The predictions are shown for typical values of $M_{\rm cut}$. The last two lines give the values of the  effective couplings ${\overline{\alpha}_s} (\overline{\mu}^{\rm PMC}_r)$ and the underlying effective scale $\overline{\mu}^{\rm PMC}_r$, respectively. The initial scale is taken as  $\mu_r=m_t$. } \label{tab_mtcut}
\end{table}
\end{center}
\end{widetext}

\begin{figure}[htb]
\includegraphics[width=0.48\textwidth]{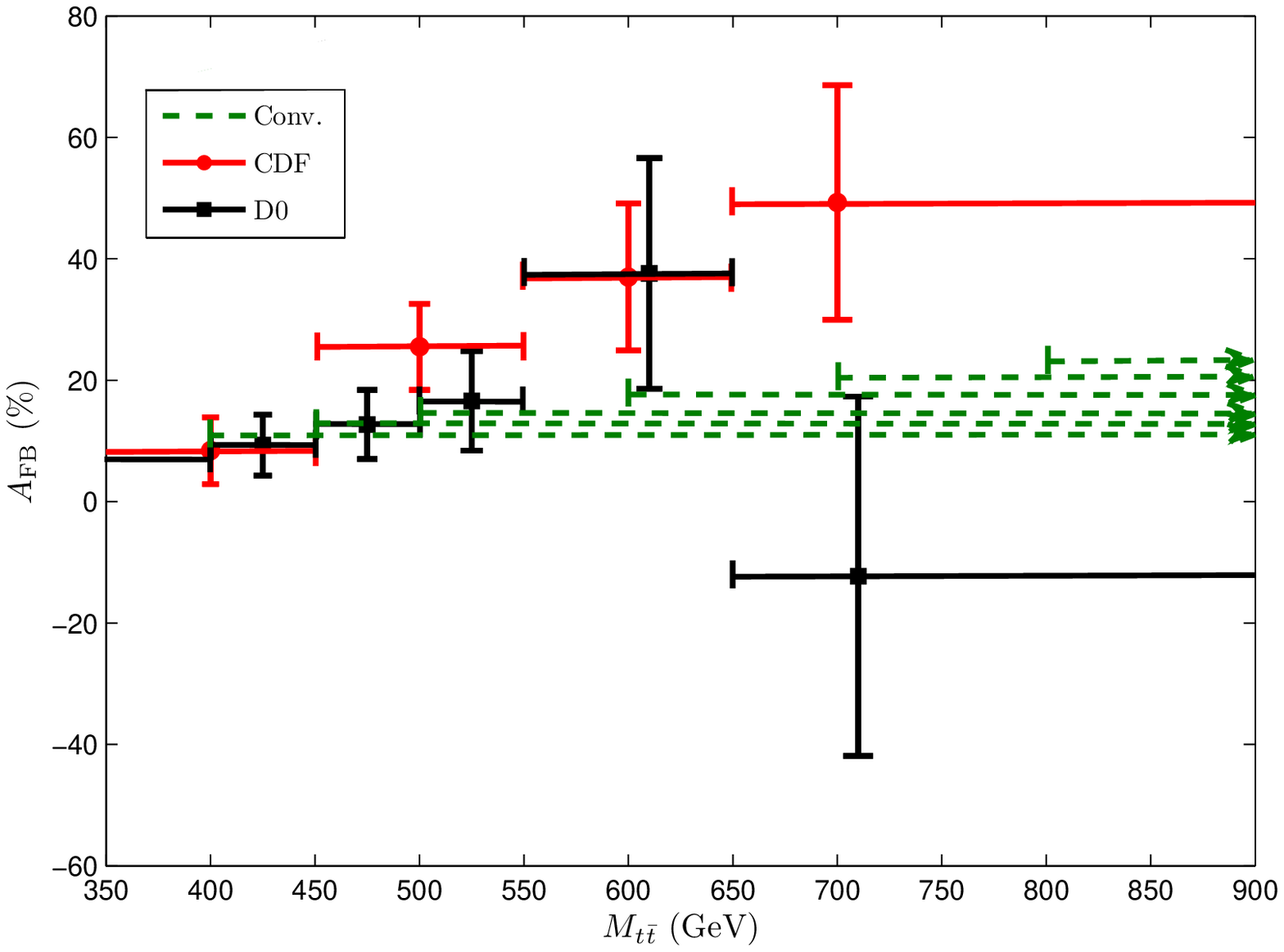}
\includegraphics[width=0.48\textwidth]{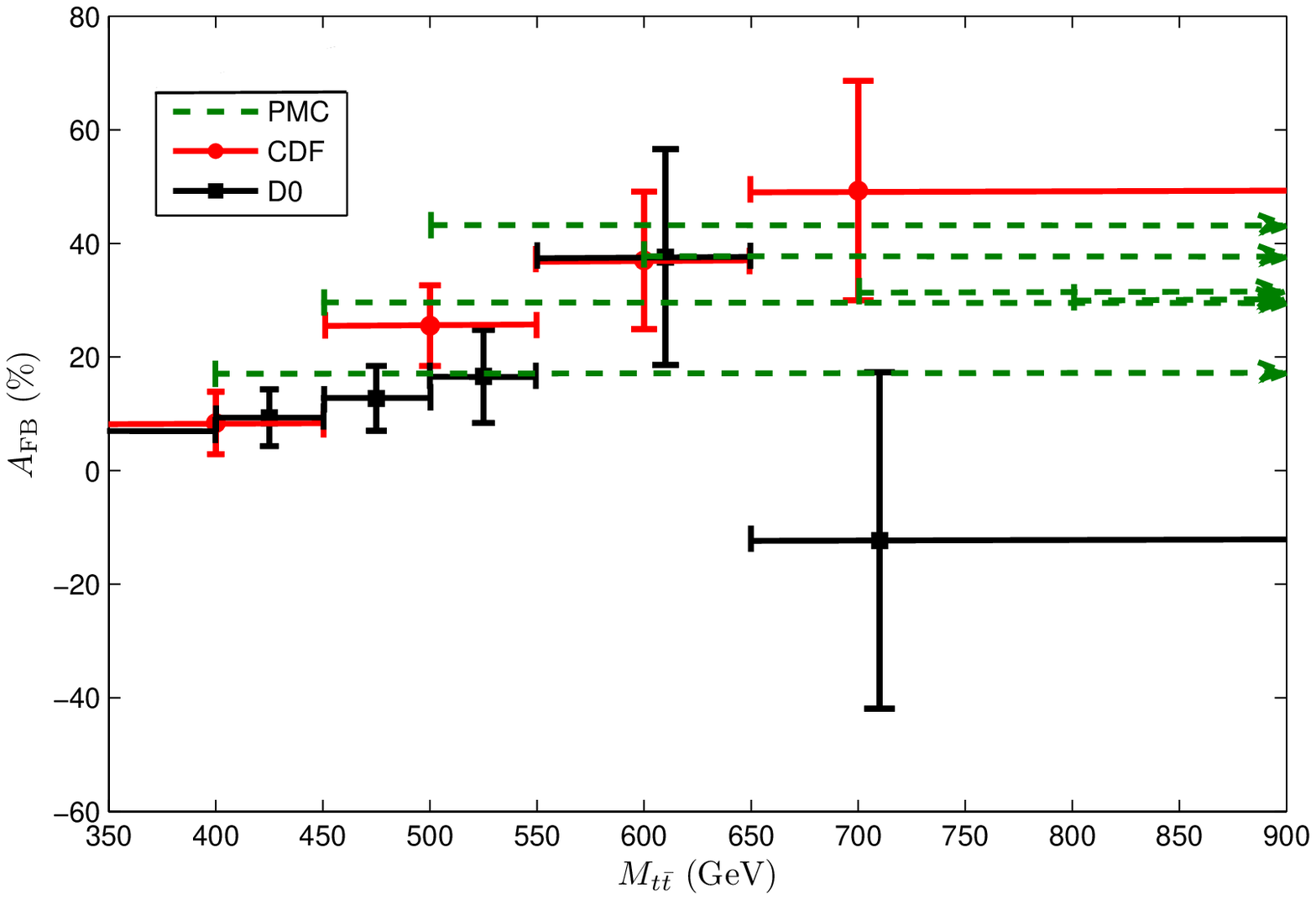}
\caption{A comparison of SM predictions of $A_{\rm FB}$ using conventional (Conv.) and PMC scale-settings with the CDF~\cite{Aaltonen:2012it} and D0~\cite{Abazov:2014cca} measurements. The Conv. predictions are for the NLO pQCD predictions with $\mathcal{O}(\alpha^2_s\alpha)$ and the $\mathcal{O}(\alpha^2)$ electroweak contributions and the PMC predictions are calculated by Eq.(\ref{pmcasy}). The upper diagram is for conventional scale-setting, and the lower one is for PMC scale-setting. The initial scale is taken as $\mu_r=m_t$. } \label{fpmcafbetw}
\end{figure}

At present, the differential $A_{\rm FB}(M_{t\bar{t}})$ is not available for us, we thus cannot directly compare with the results for $A_{\rm FB}(M_{t\bar{t}})$ given in the literature, and instead, we give the asymmetry $A_{\rm FB}(M_{t\bar{t}}>M_{\rm cut})$ for several choices of minimum pair invariant mass. The results are given in Table~\ref{tab_mtcut}, where the conventional results are for the NLO pQCD predictions with the electroweak corrections calculated by using the Bernreuther-Si program~\cite{Bernreuther:2012sx}. This behavior reflects the running behavior of the QCD coupling $\alpha_s(Q^2)$ for different kinematic regions. If one assumes conventional scale-setting and the fixed scale $m_t$, then $A_{\rm FB}(M_{t\bar{t}}>M_{\rm cut})$ monotonically increases with increasing $M_{\rm cut}$. This trend is consistent with the behavior of the N$^2$LO differential $A_{\rm FB}(M_{t\bar{t}})$ with the conventional scale-setting~\cite{Czakon:2014xsa}. In contrast, if one employs the PMC scale-setting, then $A_{\rm FB}(M_{t\bar{t}}>M_{\rm cut})$ first increases and then decreases as the lower limit of the pair mass $M_{\rm cut}$ is increased. These trends are more clearly shown by Fig.(\ref{fpmcafbetw}), in which the Standard Model predictions using conventional and PMC scale-settings are compared with the CDF~\cite{Aaltonen:2012it} and D0~\cite{Abazov:2014cca} measurements.

This ``increasing-decreasing" behavior can be understood in terms of the effective pQCD coupling $\bar\alpha_s(\overline{\mu}^{\rm PMC}_r)$ introduced in Ref.\cite{Brodsky:2012ik}; it is the weighted average of the running couplings entering the $(q\bar{q})$-channel, the subprocess underlying the asymmetry in pQCD. The effective coupling can be unambiguously determined because the NLO-level asymmetric contribution from $(q\bar{q})$-channel only involves a single PMC scale. The effective coupling $\bar\alpha_s(\overline{\mu}^{\rm PMC}_r)$ depends in detail on the kinematics. More explicitly, as shown by Table~\ref{tab_mtcut}, the non-monotonic behavior of the effective coupling accounts for the ``increasing -decreasing" behavior of $A_{\rm FB}(M_{t\bar{t}}>M_{\rm cut})$. We have adopted the partonic center-of-mass frame to estimate the PMC total cross-section for various $M_{t\bar{t}}$ cuts; the results, however, agree with the results obtained that of $t\bar{t}$-rest frame within high accuracy since events near the partonic threshold give the dominant contribution to the cross-section at the Tevatron~\cite{Ahrens:2011mw, Ahrens:2011uf}.

\section{Conclusions}

In summary, we have shown that the PMC provides a systematic and unambiguous way to set the optimal renormalization pQCD scales for top-quark pair production at each order of $\alpha_s$. The resulting pQCD predictions are renormalization-scheme independent, since all of the scheme-dependent $\{\beta_i\}$-terms in the QCD perturbative series are resummed into the QCD running couplings at each order. By applying the PMC, one obtains not only scheme-independent predictions but also a more convergent pQCD series without factorial renormalon divergences.

After applying the PMC, the uncertainties from renormalization scale-setting for both the total cross-section and the individual cross-sections at each order are simultaneously eliminated, and the residual scale dependence due to unknown higher-order $\{\beta_i\}$-terms is found to be negligible. After applying the PMC, we improved the prediction respect to renormalization scale dependence and obtain the top-quark pair forward-backward asymmetry $A^{\rm PMC}_{\rm FB}=9.2\%$ and $A_{\rm FB}(M_{t\bar{t}}>450\;{\rm GeV})=29.9\%$, in agreement with the corresponding CDF and D0 measurements.

The PMC prediction for the top-quark pair forward-backward asymmetry $A_{\rm FB}(M_{t\bar{t}}>M_{\rm cut})$ displays an ``increasing-decreasing" behavior as $M_{\rm cut}$ is increased. This behavior however cannot be explained even by a N$^2$LO QCD calculation using conventional scale-setting, since in contrast, it predicts a monotonically increasing behavior. We have also shown in a recent paper~\cite{Wang:2014sua} that the PMC predictions are in agreement with the available ATLAS and CMS data.

Thus, the proper setting of the renormalization scale provides a consistent Standard Model explanation of the top-quark pair asymmetry measurements at both the Tevatron and LHC. It is noted that the large discrepancy between the conventional lower-order prediction and the data can be cured to a certain degree by including high-order terms, such as the state-of-art results of Refs.\cite{Czakon:2014xsa, Kidonakis:2015ona}, however in those works the renormalization scale uncertainty is ``suppressed" but not ``solved". The PMC results demonstrate that the application of the PMC eliminates a major theoretical uncertainty for pQCD predictions, thus increasing the sensitivity of the LHC and other colliders to possible new physics beyond the Standard Model.  \\

\noindent{\bf Acknowledgments}: This work was supported in part by the Natural Science Foundation of China under Grant No.11275280, No.11325525, No.11547010 and No.11547305, the Department of Energy Contract No.DE-AC02-76SF00515, and by Fundamental Research Funds for the Central Universities under Grant No.CDJZR305513. SLAC-PUB-16368.

\end{document}